# About Summarization in Large Fuzzy Databases

Ines Benali-Sougui
Université Tunis El Manar
Ecole Nationale d'Ingénieurs de Tunis
BP. 37, Le Belvédère 1002 Tunis, Tunisia
ines.benali@gmail.com

Minyar Sassi-Hidri
Université Tunis El Manar
Ecole Nationale d'Ingénieurs de Tunis
BP. 37, Le Belvédère 1002 Tunis, Tunisia
minyar.sassi@enit.rnu.tn

Amel Grissa-Touzi
Université Tunis El Manar
Ecole Nationale d'Ingénieurs de Tunis
BP. 37, Le Belvédère 1002 Tunis, Tunisia
amel.touzi@enit.rnu.tn

*Abstract*—Moved by the need increased for modeling of the fuzzy data, the success of the systems of exact generation of summary of data, we propose in this paper, a new approach of generation of summary from fuzzy data called "Fuzzy-SaintEtiQ". This approach is an extension of the SaintEtiQ model to support the fuzzy data. It presents the following optimizations such as 1) the minimization of the expert risk; 2) the construction of a more detailed and more precise summaries hierarchy, and 3) the co-operation with the user by giving him fuzzy summaries in different hierarchical levels.

## I. INTRODUCTION

In the field of the Databases (DB), volumes of the data reached today make necessary a better exploitation of the data.

Several approches have been proposed to solve this problem and to contribute in database summaruzation. However, to support massive data evolutionary, formal approaches have been proposed to surround this problem. [1][2][3][4].

Several methods of DB summarization have been proposed such as statistical approaches, classification and conceptual classification. Among the methods of generation of summary of data, which is close to our research tasks, we distinguish the system SaintEtiQ [1] which is inspired primarily by the approach of conceptual classification. This system makes it possible to generate a hierarchy of summaries making it possible to cover parts of the data base.

In addition with the evolution of the data processing, the need for modeled fuzzy data became a necessitate of the user. Indeed, in the real world, we are confronted more and more with the situation where applications need to manage fuzzy data and to make profit their users from flexible querying.

Users want to express query preferences and thus obtain approximate answers [5].

We speak then about flexible querying and Fuzzy Data Bases (FDB) [6, 7, 8].

In this paper, we propose an extension of the SaintEtiQ summarization model for modeling fuzzy data and by presenting some optimization: 1) the minimization of the risk of the expert domain; 2) the construction of an hierarchy of summaries more detailed and more precise, and 3) the co-operation with the user by giving him summaries in different levels from the hierarchy. This approach is based on the combination of fuzzy logic, fuzzy clustering and Formal Concept Analysis (FCA).

The rest of this paper is organized as follows: Section 2 presents an overview of some summarization model and the basic concepts of Fuzzy Database. Section 3 presents an example of fuzzy data. Section 4 presents problems and limits of the existing summarization approach. Section 5 presents our Fuzzy-SaintEtiq system which we propose. Section 6 presents a comparison between our summary model Fuzzy-SaintEtiq and the other models. We finish this paper with a conclusion and a presentation of some future works.

## II. BASIC CONCEPTS

In this section, we present an overview of some summarization model and the basic concepts of Fuzzy Database.

### A. Overview of the SaintEtiQ summarization model

The SaintEtiQ model [1] aims at apprehending the information from a DB in a synthetic manner. This is done through linguistic summaries structured is a hierarchy. The model offers different granularities, i.e. levels of abstraction, over the data. The system architecture and the steps necessary to build a hierarchy are described below. With SaintEtiQ model, the summarization process can be divided into three major steps shown on Figure 1.

Figure 1. The Overall process of of DB summarization in SaintEtiQ model[1].

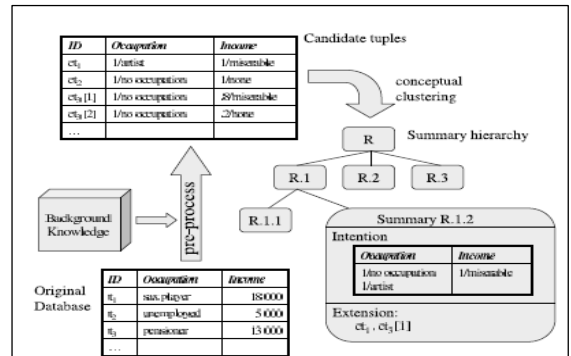

- **A Translation step**: this step allows the system to rewrite DB records in order to be processed by the mining algorithm. This translation step gives birth to candidate records, which are different representations of a single DB record, according to some background knowledge. Background knowledge's are fuzzy partitions defined over attribute domains. Each class of a partition is also labeled with a linguistic descriptor provided by the user or a domain expert. For instance, the fuzzy label young could belongs to a partition built over the domain of the attribute AGE.
- **A data mining step**: it considers the candidate records one at a time, and performs a scalable machine learning algorithm to extract knowledge. Obviously, the intensive use of background knowledge, which supports the translation step, avoids finding surprising knowledge nuggets.
- **A post processing step**: SaintEtiQ model tries to define summaries at different level of granularity. The post-processing step consists in organizing the extracted summaries into a hierarchy, such that the most general summary is placed at the root of the tree, and the most specific summaries are the leaves.

### B. Overview of FCA-based Summary

In [5], we have proposed to extend the SaintEtiQ summarization model [1] by introducing some optimization processes including: i) minimization of the expert risks domain, iii) building of the summary hierarchy from DB records, and iv) cooperation with the user by giving him summaries in different hierarchy levels. With our model, the summarization process can be divided into two major phases as shown on Figure 2.

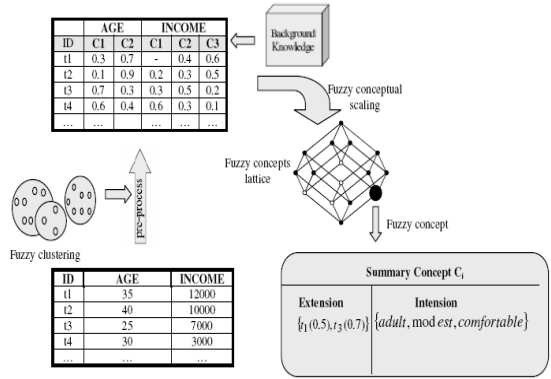

Figure 2. The Overall process of proposed FCA-based summary model[5]

### C. Fuzzy Database

In this section, we present the basic concepts of Fuzzy Database.

A Fuzzy Database (FDB) is an extension of the relational database. This extension introduces fuzzy predicates under shapes of linguistic expressions that, at the time of a flexible querying, permits to have a range of answers (each one with a membership degree) in order to offer to the user all intermediate variations between the completely satisfactory answers and those completely dissatisfactory [8].

The FRDB models are considered in a very simple shape and consist in adding a degree, usually in the interval [0,1], to every tuple.

It allows maintaining the homogeneity of the data in DB. The main models are those of Prade-Testemale, Umano-Fukami, Buckles-Petry, Zemankova-Kaendel and GEFRED of Medina et al. [9].

This last model constitutes an eclectic synthesis of the various models published so far with the aim of dealing with the problem of representation and treatment of fuzzy information by using relational DB.

### D. The GEFRED Model

The GEFRED model (GEneralised model Fuzzy heart Relational Database) has been proposed in 1994 by Medina et al. [9].

One of the major advantAGEs of this model is that it consists of a general abstraction that allows for the use of various approaches, regardless of how different they might look.

In fact, it is based on the generalized fuzzy domain and the generalized fuzzy relation, which include respectively classic domains and classic relations.

In order to model fuzzy attributes we distinguish between two classes of fuzzy attributes: Fuzzy attributes whose fuzzy values are fuzzy sets and fuzzy attributes whose values are fuzzy degrees [6, 10].

**Fuzzy Sets as Fuzzy Values**: These fuzzy attributes may be classified in four data types. This classification is performed taking into account the type of referential or underlying domain. In all of them the values Unknown, Undefined, and Null are included:
- **Fuzzy Attributes Type 1 (FTYPE1)**: These are attributes with "precise data", classic or crisp (traditional, with no imprecision). However, they can have linguistic labels defined over them, which allow us to make the query conditions for these attributes more flexible.
- **Fuzzy Attributes Type 2 (FTYPE2)**: These attributes admit both crisp and fuzzy data, in the form of possibility distributions over an underlying ordered domain (fuzzy sets). It is an extension of the FTYPE1 that does, now, allow the storage of imprecise information.
- **Fuzzy Attributes Type 3 (FTYPE3)**: They are attributes over "data of discreet non-ordered dominion with analogy". In these attributes some labels are defined ("blond", "red", "brown", etc.) that are scalars with a similarity (or proximity) relationship defined over them, so that this relationship indicates to what extent each pair of labels be similar to each other.
- **Fuzzy Attributes Type 4 (FTYPE4)**: These attributes are defined in the same way as Type 3 attributes, without it being necessary for a similarity relationship to exist between the labels.

**Fuzzy Degrees as Fuzzy Values**: The domain of these degrees can be found in the interval [0,1], although other values are also permitted, such as a possibility distribution (usually over this unit interval) [9,10]. The meaning of these degrees is varied and depends on their use. The most important possible meanings of the degrees used by some authors are: Fulfillment degree, Uncertainty degree, possibility degree and Importance degree.

*E. The FSQL language*

The FSQL language is an authentic extension of SQL language to model fuzzy queries. It means that all the valid statements in SQL are also valid in FSQL [9, 10].

## III. EXAMPLE OF FUZZY DATA

In this example, we want to model an employee described by the following information: his *Id* (identifier), his name, his surname, his address, his *Age*, his *Salary*, and his productivity. Attributes *Age*, *Salary* and *Productivity* are described as follows:

- The attribute *Age*, presented in Figure 3, has the linguistic labels *Young*, *Adult* and *Old*, defined on the trapezoidal possibility distributions as following: *Young(18, 22, 30, 35), Adult(25, 32, 45, 50), Old(50, 55, 62, 70)*. An approximate value has a margin of 5. The minimal value to consider two values of this attribute as completely different is of 10.

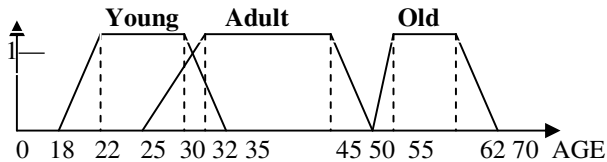

Figure 3. Definition of labels on Age attribute

- The attribute salary, presented in Figure 4, has the linguistic labels *Low*, *Medium* and *High*, defined on the trapezoidal possibility distributions as following: *Low(50,80,120,180), Medium(150,300,400,550), High(400, 600,800,1000)*. An approximate value has a margin of 10 and the minimal value to consider two values of this attribute as completely different is of 50.

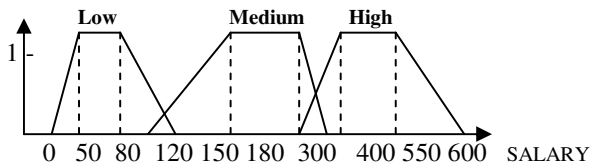

Figure 4. Definition of salary labels

- The attribute *productivity*, presented in Table I, has the linguistic labels *Bad, Regular* and *Good*. In this situation, the data are not quantifiable, but present resemblances in their values. For example, the value *Regular* of the attribute "productivity" resembles to the value *Good* with a degree equal to 0.7.

TABLE I. RELATIONS OF SIMILARITY FOR THE VALUES OF THE PRODUCTIVITY ATTRIBUTE

| Similarity degree | BAD | REGULAR | GOOD |
|---|---|---|---|
| BAD | 1 | 0.3 | 0.2 |
| REGULAR | 0.3 | 1 | 0.7 |
| GOOD | 0.2 | 0.7 | 1 |

While applying the rules of Medina et al., we can say that the *AGE* attribute is of FTYPE2 (5, 10) type, the attribute *salary* is of FTYPE1(10,50) type and the attribute *productivity* is of FTYPE3(1) type.

An abstract representation of the schema of relation EMPLOYE will be as follows: (ID, NAME, SURNAME, ADDRESS, *AGE, SALARY, PRODUCTIVITY*). This description in FSQL script is presented in the Figure 5.

```
CREATE TABLE EMPLOYEE (
ID# VARCHAR(4) NOT NULL,
NAME VARCHAR(20) NOT NULL,
SURNAME VARCHAR(20) NOT NULL,
ADDRESS VARCHAR(40) NOT NULL,
AGE FTYPE2(5,10) NUMBER(3) DEFAULT UNKNOWN
NOT NULL,
SALARY FTYPE1(10,50) NUMBER(7) NOT NULL,
PRODUCTIVITY FTYPE3(1) NOT NULL,
PRIMARY KEY (ID#));
```

Figure 5. FSQL script

*A. Problems and Motivation*

We present in the following table a synthesis of the existing summarization techniques. As Table II depicts, these approach are applicable only to simple data sets. And don't permit to treat fuzzy data, describe with FSQL language, like linguistic labels (string), interval, and approximate values.

In this paper, we propose to define a new approach of summarization allowing treating as well the simple data set or the fuzzy data describe with FSQL language.

TABLE II. COMPARATIVE STUDY OF SOME SUMMARIZATION TECHNIQUES

|  | Understandable | Sampling | Data Nature | Huge data | Ratio with the original data | hierarchical levels | Reliability | Subject depending | Fuzzy DB |
|---|---|---|---|---|---|---|---|---|---|
| Statistical Model | comprehensible | No | Numeric/ Nominal | No | Lost | No | High | Yes | No |
| Classification | No | Yes | Numeric | No | Kept | No | Low | No | No |
| Conceptual classification | Partially | Yes | Numeric | No | Kept | No | Means | No | No |
| SaintEtiQ | Yes | Yes | Numeric/ Nominal | Yes | Kept | Partially | Means | No | No |
| FCA-based Summary [5] | Yes | Yes | Numeric | Yes | Kept | Yes | High | No | No |

## IV. MODEL DESCRIPTION

In this section, we present the architecture of the summarization model, the principal of summaries generation and the formal description of summary.

### A. System architecture

Our summary model takes the database records and provides knowledge.

Figure 6 gives the system architecture. The summarization act considered like a process of knowledge discovery from database, in the sense that it is organized according to two following principal steps.

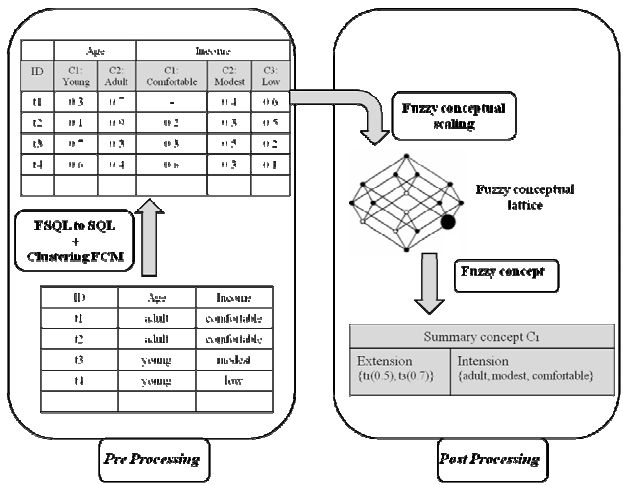

Figure 6. The Overall process of Fuzzy-SaintEtiq

#### 1) The preprocessing step

This step organizes the database records in homogeneous clusters having common properties. This step gives a certain number of clusters for each attribute. Each tuple has values in the interval [0..1] representing these membership degrees according the formed clusters. Linguistic labels, which are fuzzy partitions, will be attributed on attribute's domain.

The classification on these fuzzy data uses the Fuzzy FCM algorithm. Fuzzy-FCM is an extension FCM algorithm in order to support different types of data represented by GEFRED model. Figure 7 shows the different steps of this algorithm.

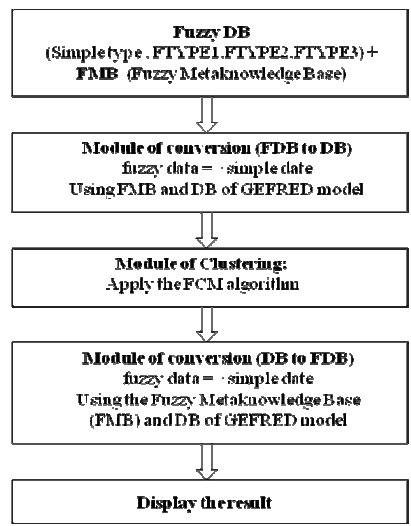

Figure 7. Principe of FCM-FDB algorithm

The Fuzzy-FCM algorithm allows the user to select attributes according to which he wants to carry out classification, which gives a refined intermediate matrix only formed of the codes of the selected attributes.

Once the selection achieved the FCM algorithm is applied on the refined table to get a matrix of adherence and a cut is exercised on this matrix of adherence to purify it that is to eliminate all values lower to the cut.

The main idea of the algorithm is to define an intermediate matrix to model the fuzzy data set to model the module of conversion. For this, we define the function $\mathcal{F}$ permit to construction this matrix. F is defined as follows:

**Definition 1**: Let E the set of linguistic labels and C the set of numbers.

We define $\mathcal{F}$ as a function which for all e belonging to E, makes correspond a code c belonging to C the set of correspondence codes:

$\mathcal{F}$ : E $\longrightarrow$ C
e $\longrightarrow$ c = Number of Attribute.Threshold

Since the attributes of the type FTYPE1 do not authorize to store fuzzy values they undergo the same treatment as the simple data and thereafter the function $\mathcal{F}$ = id.

For the attributes of the type FTYPE 2 and FTYPE 3, the function $\mathcal{F}$ makes correspond to each linguistic label a code of the form NumberAttribut.Threshold.

We define the Threshold as being the minimal value to be able considers two values as completely different.

**Example:** Let's consider the relational DB table Personal, represented in Table III, described by Id, Age, and Experience.

The attribute Age has the linguistic labels definite on the following trapezoidal distributions possibility : Young (18,22,30,35), Adult (25, 32,45,50), Old(50,55,62,70). The minimal value to consider two values of this attribute as completely different is 10.

The attribute Experience has the linguistic labels: Small(2,3,5,6), Good(5,7,10,12), Sufficient(7, 8,15,20), Large(12,15,50,50). These values depend on the numbers of years worked by an employee. The minimal value to consider two completely different Experiences as is 5.

Given the Personal DB table represented in Table III.

TABLE III.  PERSONAL DB TABLE

| Id | Age | Experience |
|---|---|---|
| 001 | Young | Good |
| 002 | Old | Small |
| 003 | Adult | Sufficient |
| 004 | Young | Large |

While applying the rules of Medina et al, we can say that the AGE attributes and Experience attribute is FTYPE2. Thus, the correspondence table is presented by Table IV.

TABLE IV.  TABLE OF CORRESPONDANCE

| Id | Age | Experience |
|---|---|---|
| 001 | 1.1 | 2.15 |
| 002 | 1.3 | 2.10 |
| 003 | 1.2 | 2.20 |
| 004 | 1.1 | 2.25 |

For the first attribute Age of this table, the choice of the number 1 translated the number of the attribute on which one works. Here the Age attribute is the attribute number 1 and thereafter codes corresponding to the linguistic labels start all with 1. The minimal value to consider two values of this attribute as completely different is 10, we fix a step then = 10 in the choice of codes. For the second attribute experience, the choice of the number 2 translated the number of the attribute on which one works. Here the attribute Experience is the attribute number 2 and thereafter codes corresponding to the linguistic labels start all with 2. The minimal value to consider two values of this attribute as completely different is 5, we fix a step then = 5 in the choice of codes. Moreover the choice of these codes concord with the semantics of labels.

For example the small label is nearer "semantically" to the good label than of the big label, thus we chose the codes according to this logic "of ascending order".

*2) The post treatment step*

This step takes into account the result of the fuzzy clustering on each attribute, visualizes by using the fuzzy concepts lattices. Then, it imbricates them in a fuzzy nested lattice.

Finally, it generalizes them in a fuzzy lattice associating all records in a simple and hierarchical structure. Each lattice node is a fuzzy concept which represents a concept summary.

This structure defines summaries at various hierarchical levels.

This step consists in organizing the summaries within a hierarchy such that the most general concept summary is placed at the root of the fuzzy lattice, and the most specific concept's summaries are the leaves.

This summary model corresponds to prototypical approaches since the intention of a concept summary present for each attribute the various possible values in the form of a fuzzy descriptors and the representativeness of these descriptors within the specified concept summary.

This model will be described formally in subsection C.

*B. Principal of summaries generation*

The summary model presented here is based on the fuzzy subsets theory with each one of its steps.

*1) Generating attribute's clusters*

For the generation of the clusters for each attribute, we carry out a fuzzy clustering while benefiting from fuzzy logic. This operation makes it possible to generate, for each attribute, a set of membership degrees. Each cluster of a partition is labeled by linguistic descriptor provided by a domain expert.

For example, the fuzzy label young belongs to a partition built on the domain of attribute AGE.

*2) Building the summary hierarchy*

After the generation of the clusters of each attribute, data are ready to be summarized. This operation is based on the fuzzy lattices notion.

This very simple sorting procedure gives us for each many-valued attribute the distribution of the objects in the line diagram of the chosen fuzzy scale. Usually, we are interested in the interaction between two or more fuzzy many-valued attributes. This interaction can be visualized using the so-called fuzzy nested line diagrams. It is used for visualizing larger fuzzy concept lattices, and combining fuzzy conceptual scales on-line.

From the fuzzy nested lattice, we can draw a summary hierarchy of the same fuzzy context. This illustrated in Figure 8.

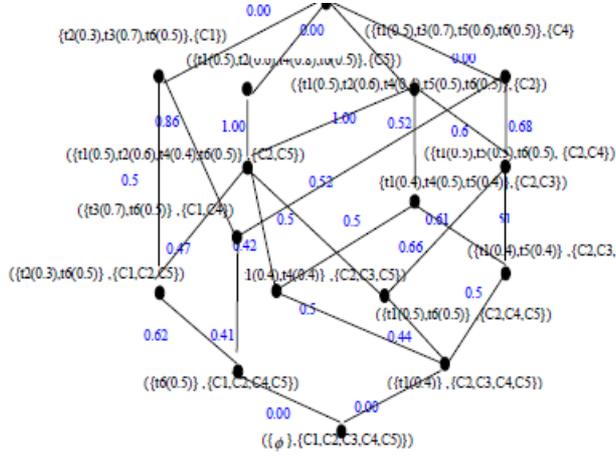

Figure 8. Principal of FCM-FDB algorithm

### C. Formal representation of summaries

Such as shown in Figure 8, each concept summary can be viewed like an n-uplet of a relation $R^*$ whose diagram is the same one as the origin relation $R$ to summarize. Each concept summary $z$ of the set of concept's summary $Z$ is thus a description of a set of n-uplets of $R$, which jointly form his extension and which is noted by $R_z$.

**Definition 2. Concept summary**: A concept summary is a couple $z = (R_z, I_z)$ in which $R_z$ is the subset of database records involved into the summarization, the extent, whereas the summarized description $I_z$ of these database records is the intent.

Each concept summary $z = (R_z, I_z)$ provides a synthetic view of a part of the database.

Thus, the root contains the summary of all the candidate records, whereas leaves represent only one combination of fuzzy linguistic labels over all the attributes.

**Example:** $z = (\{t1(0.5), t5(0.5), t6(0.5), \{modest, young\})$

**Definition 3. Abstraction level**: an abstraction level is regarded as a level in the summary hierarchy generated.

**Definition 4. Level**: A level $L$ of a summary hierarchy is a set of concept's summary $z_k$ verifying the following property: the majors and the minors of $z_k$ are at the same distance $d$.

**Definition 5. Majors/Minors**: Let $(E, \leq_E)$ be an ordered set and $S$ a subset of $E$. Major's elements (successors) and Minor's elements (predecessors) of $S$ are defined by:

$$\text{Majors}(S) = \{x \in E \land \forall y \in S, y \leq_E x\}$$
$$\text{Minors}(S) = \{x \in E \land \forall y \in S, x \leq_E y\}$$

Considering the summary hierarchy in Figure 8, we can generate the following levels with the corresponding summaries:

**Level 0** { $z1 = (\{t1(0.0), t2(0.0), t3(0.0), t4(0.0), t5(0.0), t6(0.0)\}, \{\phi\})$

**Level 1**
- $z21 = \{t2(0.3), t3(0.7), t6(0.5)\}, \{\text{miserable}\})$
- $z22 = t1(0.5), t2(0.6), t4(0.8), t6(0.5)\}, \{\text{adult}\})$
- $z23 = (\{t1(0.5), t2(0.6), t4(0.4), t5(0.5), t6(0.5)\}, \{\text{modest}\})$
- $z24 = (\{t1(0.5), t3(0.7), t5(0.6), t6(0.5)\}, \{\text{young}\})$

**Level 2**
- $z31 = (\{t1(0.5), t2(0.6), t4(0.4), t6(0.5)\}, \{\text{modest, adult}\})$
- $z32 = (\{t1(0.5), t5(0.5), t6(0.5), \{\text{modest, young}\})$
- $z33 = \{t1(0.4), t4(0.5), t5(0.4)\}, \{\text{modest, comfortable}\})$
- $z34 = (\{t3(0.7), t6(0.5)\}, \{\text{miserable, young}\})$

**Level 3**
- $z41 = (\{t1(0.4), t5(0.4)\}, \{\text{modest, comfortable, young}\})$
- $z42 = (\{t1(0.4), t4(0.4)\}, \{\text{modest, comfortable, adult}\})$
- $z43 = (\{t2(0.3), t6(0.5)\}, \{\text{miserable, modest, adult}\})$
- $z44 = (\{t1(0.5), t6(0.5)\}, \{\text{modest, young, adult}\})$

**Level 4**
- $z51 = (\{t6(0.5)\}, \{\text{miserable, modest, young, adult}\})$
- $z52 = (\{t1(0.4)\}, \{\text{modest, comfortable, young, adult}\})$

**Level 5** { $z61 = (\{\phi\}, \{\text{miserable, modest, comfortable, young, adult}\})$

Levels 0 and 6 are both the root and leaves concept summary.

A concept summary is defined in an extensional manner with a collection of candidate records $R_z = \{t_1, t_2, \ldots, t_N\}$.

Each $t_i$ is associated to one primitive database records, i.e. an element of $R$.

Denote by $card(R_z) = \sum_{t \in R_z} w(t)$ the representativity of the concept summary $z$ according to the primary database $R$. $|R_z|$ the number of candidate records in $R_z$.

### D. About complexity

The space complexity, whatever the number of database records, is thus reduced to a constant value, i.e., about O(1). This characteristic is fundamental in the treatment of the large database in knowledge discovery. Temporal complexity includes the following costs:

- Construction of the attribute's clusters.
- Building the fuzzy lattice.

For cluster's construction, the complexity of fuzzy clustering algorithms is about $O(NC^2)$, where N corresponds to database table records number and C is the maximum number of clusters.

For fuzzy lattice construction, temporal complexity of lattice construction algorithm is about $O(N^2)$.

## V. COMPARATIVE STUDY

Table V gives a comparison between our summary model Fuzzy-SaintEtiq and the other models.

TABLE V. COMPARATIVE STUDY OF SOME SUMMARIZATION TECHNIQUES

|  | Understandable | Sampling | Data Nature | Huge data | Ratio with the original data | hierarchical levels | Reliability | Subject depending | Fuzzy DB |
|---|---|---|---|---|---|---|---|---|---|
| Statistical Model | comprehensible | No | Numeric/Nominal | No | Lost | No | High | Yes | No |
| Classification | No | Yes | Numeric | No | Kept | No | Low | No | No |
| Conceptual classification | Partially | Yes | Numeric | No | Kept | No | Means | No | No |
| SaintEtiQ[1] | Yes | Yes | Numeric/Nominal | Yes | Kept | Partially | Means | No | No |
| FCA-based Summary [5] | Yes | Yes | Numeric | Yes | Kept | Yes | High | No | No |
| Fuzzy_SaintEtiq | Yes | Yes | Numeric | Yes | Kept | Yes | High | No | Yes |

## VI. CONCLUSION

In the last decades, Fuzzy FSQL language had a big success for the description and the manipulation of the BDRFS. In this paper, we proposed a new approach to linguistic summarization for fuzzy databases, called fuzzy-SaintEtiq.

To validate our approach, we currently plan to develop this approach with JAVA language.

As futures perspectives of this work, we mention essentially 1) to test our approach on the large fuzzy data set and 2) to describe a new approach for Knowledge Discovery in Fuzzy Databases (KDFD) described with FSQL language.

## REFERENCES

[1] G. Raschia and N. Mouaddib, SaintEtiQ: A fuzzy set-based approach to database summarization. Int. Journal of Fuzzy Sets and Systems, 129(2), pp. 137–162, 2002.

[2] RR. Yager, A new approach to the summarization of data. Information Sciences, 28(1), pp. 69–86, October 1982.

[3] P. Bosc, O. Pivert and L. Ughetto, On data summaries based on gradual rules, Computational intelligence : theory an applications, pp. 512–521, 1999.

[4] H.J. Lenz and A. Shoshani, Summarizability in OLAP and statistical data bases. In SSDBM '97: Proceedings of the Ninth International Conference on Scientific and Statistical Database Management, pp. 132–143, 1997.

[5] M.Sassi, A. Grissa-Touzi, H. Ounelli, I. Aissa, About Database Summarization. International Journal of Uncertainty, Fuzziness and Knowledge-Based Systems 18(2), pp.133-151, 2010.

[6] J.Galindo, A.Urrutia, and M. Piattini, Fuzzy databases: modeling, design and implementation. USA: Idea Group Publishing Hershey. 2006.

[7] M.A Ben Hassine, A. Grissa Touzi, J. Galindo, and H. Ounelli, "How to Achieve Fuzzy Relational Databases", in "Handbook of Research on Fuzzy Information Processing in Databases", Ed, Information Science Reference, pp 351- 380, Mai 2008.

[8] P. Bosc, L. Liétard, and O. Pivert, "Bases de données et Flexibilité : Les requêtes Graduelles", Techniques et Sciences informatiques, vol. 7, no. 3, pp. 355-378, 1998.

[9] J.M. Medina, O. Pons, and M.A.Vila, "GEFRED. A Generalized Model of Fuzzy Relational Data Bases", Information Sciences, vol. 76(1-2), pp. 87-109, 1994.

[10] J.Galindo, "New characteristics in FSQL, a fuzzy SQL for fuzzy databases". WSEAS Transactions on Information Science and Applications 2, 2(2), pp. 161-169, 2005.